\documentclass{PoS}

\title{Recent results from heavy flavour physics on the lattice}

\ShortTitle{Recent results from heavy flavour physics on the lattice}

\author{\speaker{Michele Della Morte}\\
       Institut f\"ur Kernphysik and Helmholtz Institut, University of Mainz,\\
       Johann-Joachim-Becher Weg 45, D-55099 Mainz, Germany\\
       E-mail: \email{morte@kph.uni-mainz.de}}

\abstract{
\vspace{-8.4cm}
We discuss  some recent lattice results on a few selected hadronic
 quantities relevant for heavy flavour phenomenology and present some recent 
theoretical developments. We 
put the emphasis on the challenges, which have to be faced, on the way 
to precise heavy flavour physics from the lattice. We also discuss the 
importance in the search for New Physics of possible future lattice
studies of the form factors entering $B~\to~K$ and $B~\to~K^*$ 
semileptonic decays.
\vspace{-13.9cm}
\begin{flushright}
MKPH-T-10-36\\
\end{flushright}
}

\FullConference{35th International Conference of High Energy Physics - ICHEP2010,\\
		July 22-28, 2010\\
		Paris France}

\begin{document}

\vspace{-0.3cm}
\section{Heavy quarks on the lattice }
\begin{figure}[htb]
\begin{center}
\includegraphics[width=10.cm]{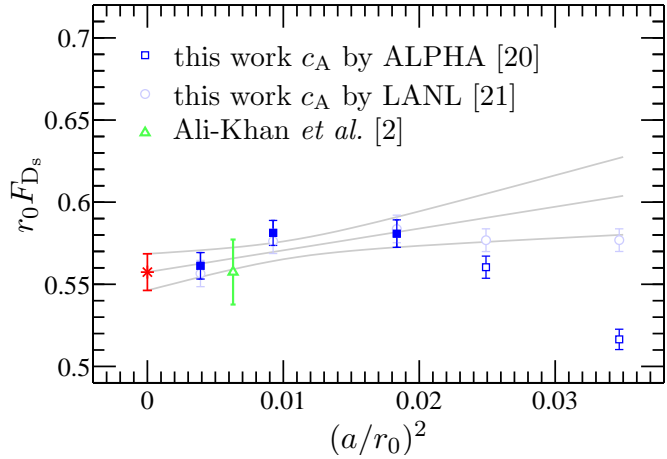}
\vspace{-0.7cm}
\caption{The (quenched) decay constant $F_{\rm D_{\rm s}}$ vs the 
lattice spacing for the O$(a)$ improved Wilson-Clover action and two
 different choices of the improved current. The lattice spacing is 
measured in units of the scale $r_0 \simeq 0.5$ fm~\cite{r0}. 
Plot from~\cite{jutt}.} 
\label{fdsA}
\end{center}
\end{figure}

\vspace{-0.3cm}
Heavy flavour phenomenology is an important tool for the indirect search of 
New Physics (NP) and for constraining possible extensions of the Standard 
Model (SM). As these indirect NP effects are expected to be small~\cite{Artuso}, it is
absolutely necessary to have all systematics affecting the theoretical 
predictions under control. In this framework the lattice can provide first
principle determinations of important quantities such as heavy mesons decay 
constants and form factors for rare B-decays. Cutoff effects, i.e. the dependence
of the results on the lattice resolution, are one of the main systematics in this
case. It is clear that in order to properly describe the propagation of 
heavy quarks fine lattice resolutions are needed. The issue has been quantitatively studied in~\cite{jutt}, by computing
the decay constant $F_{\rm D_{\rm s}}$ of the $D_{\rm s}$ meson, in the
quenched approximation,
over a large
range of lattice spacings $a$, namely $ 0.03 {\; \rm fm} \lesssim a \lesssim 0.09 {\; \rm fm}$. The results are shown in fig.~\ref{fdsA}. 
%(taken from~\cite{jutt}).
%
\vspace{-1.cm}
\vspace{-0.6cm}
As the action and the operators are both O($a$) improved, scaling 
violations
are expected to be linear in $a^2$. However such a behaviour sets in 
for small
lattice resolutions only, $a\lesssim 0.07$ fm (i.e. the three finest
 resolutions in the plot). If  results for $a \gtrsim 0.07$ fm only
had been available then the continuum extrapolation would have produced
a rather different number for $F_{\rm D_{\rm s}}$. This despite the
fact that the data show a linear dependence on $a^2$ in that region
of lattice spacings. A second set of numbers is obtained by adopting
another definition of the improved axial current 
(grey symbols in the figure), which differs from the first one by 
O($a^2$). The cutoff effects appear to be smaller for this second 
definition, however the situation is reversed  for other observables
as the charm quark mass, as discussed in~\cite{jutt}.
It is clear anyway that by considering two different O$(a$) improved
regularizations of the current some indications can be gathered
on the onset of the scaling region.
The overall lesson seems to be that Symanzik improvement programme
works also for heavy quarks, but lattice spacings smaller than $0.08$ fm
are needed.

A second example on how delicate continuum limit extrapolations are 
for quantities involving heavy quarks is obtained by revisiting the 
so called $F_{\rm D_{\rm s}}$ tension. The extremely precise result 
$F_{\rm D_{\rm s}}=241(3)$ MeV obtained in 2007 by the HPQCD 
Collaboration~\cite{fdsh}
using $2+1$ flavours of rooted staggered quarks turned out to differ by more than
three standard deviations from the experimental estimate $F_{\rm D_{\rm s}}=274(11)$
 MeV (see~\cite{lat07}), produced
by CLEO-c by analyzing tauonic $D_{\rm s}$ decays~\cite{CLEO07}.
Now both numbers have moved by one or two sigma.
The result from CLEO-c with updated statistics is $F_{\rm D_{\rm s}}=
259(6)(3)$ MeV~\cite{CLEO09}.
The HPQCD Collaboration has  updated the 2007 result by incorporating 
a new and more accurate determination of the lattice spacing and by
including results at finer lattice resolutions, covering the range
$0.044\,{\rm fm} \leq a \leq 0.15\, {\rm fm}$ (to be compared with
the range $0.09\,{\rm fm} \leq a \leq 0.15\, {\rm fm}$ used 
in~\cite{fdsh}). This yields the value $F_{\rm D_{\rm s}}=248.0(25)$ 
MeV~\cite{fdsh10}. The tension has solved and the discrepancy is below
1.5 combined standard deviations.

Having emphasized the importance of simulating at very fine lattice 
spacings to study heavy flavour physics on the lattice, let us briefly
discuss some           algorithmic  difficulties in approaching the 
continuum limit. Critical slowing down of Monte Carlo algorithms is
expected near phase transitions, as the one which defines the continuum
limit. In~\cite{slowmodes} it was reported that a critical exponent 
as large as 5 can be measured in typical Monte Carlo simulations, for
quantities like the topological charge, this implies a scaling as 
$a^{-10}$ for full QCD simulations. It was on the other hand also
 observed that other quantities such as heavy-light pseudoscalar 
correlation functions couple only weakly to the slow modes affecting
the topological charge. The possibility however remains that modes even
slower than those detected in that study may be present. A robust
statistical error estimate is therefore a delicate issue and improved
estimators have been proposed in~\cite{slowmodes}. More generally
critical slowing down provides a reason to clearly prefer few but long 
Monte Carlo chains over several short ones.

The situation is different for B-physics on the lattice. In this case
the b-quark is treated non-relativistically and coarser lattice 
resolutions than those considered for D-physics can be used.
An attractive option in this respect is Heavy Quark effective Theory 
(HQET)~\cite{EH} on the lattice. HQET provides
the correct asymptotic description of QCD correlation functions in the
limit $m_{\rm b} \!\to\! \infty$. Subleading effects are described by
higher dimensional operators whose coupling constants are formally
O($1/m_{\rm b})$ to the appropriate power. Due to the appearance
of power divergent mixings (as $a$ goes to 0), the theory must be formulated 
and renormalized non-perturbatively already at the leading (static) order~\cite{RH}.
A non-perturbative
treatment of HQET on the lattice including $1/m_{\rm b}$ corrections
has been recently put forward in~\cite{papI,papII,papIII} with
first applications on the determination of the b-quark mass, the 
mesonic spectrum in the $B_{\rm s}$ meson sector and the $B_{\rm s}$
decay constant. These studies have been performed in the quenched 
approximation but their extension to the $N_{\rm f}=2$ dynamical
case is almost completed and preliminary results are already 
available~\cite{LAT10}
\vspace{-0.4cm}
\section{$B \to K$ and $B \to K^*$ semileptonic decays}
\vspace{-0.3cm}
 These rare decays are interesting potential probes of  Physics beyond the Standard Model  since NP diagrams could  contribute
at the same level as SM ones. The branching ratios are measured at Babar, Belle and
CDF at O($10^{-7} - 10^{-6}$)~\cite{HFAG}, consistent with the SM.
These modes are also among the key measurements in the LHCb physics
programme~\cite{Perret}.
Beyond the rate, several observables
can be obtained which are sensitive to different electroweak couplings.
On the theory side most of the studies concentrated on the form factors
in the large recoil region, where QCD factorization
applies~\cite{Seidel}. However, as pointed out in~\cite{Gudrun},
the low recoil region is equally relevant in constraining NP.
This  offers the possibility for the lattice to contribute, as
the low recoil region is the one reliably accessible to lattice
calculations.
As of today only preliminary results from one  study using
moving-NRQCD exist~\cite{Wingate}. This is a very difficult 
calculation  as seven
different form factors need to be computed for the decay
$B \rightarrow K^* l^+l^-$. The confidence on lattice results 
for these form factors will improve
in the future as different approaches are considered .
\vspace{-0.2cm}
\section{Conclusions}
\vspace{-0.16cm}
The precision of lattice computations of hadronic parameters related
to heavy-flavour physics has considerably increased in the last years,
as it is required for the approach to be a tool for search of New Physics.
As the accuracy improves, systematic effects previously considered to be negligible
become relevant. Such calculations therefore remain very challenging and new ideas
and techniques are constantly being developed to keep the systematics under control.
In some cases this activity even leads to a deeper uderstaning of fundamental properties
of QCD and gauge theories~\cite{Martinflow}.

At the same time new processes are being considered for signals
of New Physics. The lattice can play an important role there
in providing a first-principle determination of the relevant form factors
and matrix elements within the Standard Model.

\vspace{0.14cm}
\noindent {\bf Acknowledgements.} I wish to thank all the members of 
ALPHA 
and in particular Rainer Sommer for the pleasant collaboration.
I am grateful to Gudrun Hiller for useful discussions.


\begin{thebibliography}{99}
\bibitem{Artuso} 
M. Artuso et al., Eur.~Phys.~J.~C57 (309) 2008.
\bibitem{jutt}
J.~Heitger and A.~J\"uttner, JHEP 0905 (2009) 101. 
\bibitem{r0}
R.~Sommer, Nucl.~Phys.~B411 (1994) 839.
\bibitem{fdsh}
E. Follana, C. T. H. Davies, G. P. Lepage and J. Shigemitsu,
Phys.~Rev.~Lett.~100 (2008) 062002.
\bibitem{lat07}
M.~Della Morte,  PoS LAT2007 (2007) 008.
\bibitem{CLEO07}
K.~M.~Ecklund et al., CLEO Collaboration, Phys.~Rev.~Lett.~100 (2008) 161801.
\bibitem{CLEO09}
P.~Naik et al., CLEO Collaboration, Phys.~Rev.~D80 (2009) 112004.
\bibitem{fdsh10}
C.~T.~H.~Davies et al., HPQCD Collaboration, arXiv:1008.4018 [hep-lat].
\bibitem{slowmodes}
S.~Schaefer, R.~Sommer and F.~Virotta, arXiv:1009.5228 [hep-lat].
\bibitem{EH} E.~Eichten and B.~Hill, Phys.~Lett.~B234 (1990) 511.
\bibitem{RH} J.~Heitger and R.~Sommer, JHEP 0402 (2004) 022..
\bibitem{papI} B.~Blossier, M.~Della Morte, N. Garron and R.~Sommer,
JHEP 1006:002, 2010.
\bibitem{papII} B.~Blossier et al., ALPHA Collaboration, JHEP 1005:074, 2010.
\bibitem{papIII}B.~Blossier et al., ALPHA Collaboration, arXiv 1006.5816 [hep-lat].
\bibitem{LAT10}
B.~Blossier et al., ALPHA Collaboration, PoS (Lattice 2010) 308.
\bibitem{Perret} P.~Perret, LHCb Collaboration, arXiv:0901.2856 [hep-ex].
\bibitem{HFAG}
E.~Barberio et al.[Heavy Flavour Averaging Group], arXiv:0808.1297 [hep-ex].
\bibitem{Seidel}
M.~Beneke, T.~Feldmann and D.~Seidel, Nucl.\ Phys.\  B {\bf  612}
(2001) 25.
\bibitem{Gudrun}
C. Bobeth, G. Hiller, D. van Dyk, JHEP 1007 (2010) 098. 
\bibitem{Wingate}
Z.~Liu et al, PoS LAT2009 (2009) 242.
\bibitem{Martinflow}
M.~L\"uscher, JHEP 1008 (2010) 071.
\end{thebibliography}
\end{document}